\documentclass[12pt,preprint]{aastex}

\def\arcsec{\hbox{$^{\prime\prime}$}}

\def\lya{Ly$-\alpha$}
\def\lyb{Ly$-\beta$}

\shorttitle{\indent \def Lyman$-\alpha$ and Lyman$-\beta$ in coronal
holes} \shortauthors{Tian et al.}

\begin{document}

\title{Hydrogen Lyman$-\alpha$ and Lyman$-\beta$ radiances and profiles in polar coronal holes}

\author{Hui Tian\altaffilmark{1,2}, Luca Teriaca\altaffilmark{2}, Werner Curdt\altaffilmark{2}, Jean-Claude Vial\altaffilmark{3}}
\altaffiltext{1}{School of earth and space sciences, Peking
University, 100871, Beijing, China; tianhui924@gmail.com}

\altaffiltext{2}{Max-Planck-Institut f\"ur Sonnensystemforschung,
37191, Katlenburg-Lindau, Germany}

\altaffiltext{3}{Institut d'Astrophysique Spatiale, Unit\'{e} Mixte,
CNRS-Universit\'{e} de Paris XI, Bat 121, 91405 Orsay, France}

\begin{abstract}
The hydrogen Lyman-alpha (\lya) plays a dominant role in the
radiative energy transport in the lower transition region, and is
important for the studies of transition-region structure as well as
solar wind origin. We investigate the \lya~profiles obtained by the
Solar Ultraviolet Measurements of Emitted Radiation spectrograph
(SUMER) on the SOHO (Solar and Heliospheric Observatory) spacecraft
in coronal holes and quiet Sun. In a subset of these observations,
also the H\,{\sc i} Lyman-$\beta$ (\lyb), Si\,{\sc iii}, and O\,{\sc
vi} lines were (quasi-) simultaneously recorded. We find that the
distances between the two peaks of \lya~profiles are larger in
coronal holes than in the quiet Sun, indicating a larger opacity in
coronal holes. This difference might result from the different
magnetic structures or the different radiation fields in the two
regions. Most of the \lyb~profiles in the coronal hole have a
stronger blue peak, in contrast to those in quiet-Sun regions.
Whilst in both regions the \lya~profiles are stronger in the blue
peak. Although the asymmetries are likely to be produced by
differential flows in the solar atmosphere, their detailed formation
processes are still unclear. The radiance ratio between \lya~and
\lyb~decreases towards the limb in the coronal hole, which might be
due to the different opacity of the two lines. We also find that the
radiance distributions of the four lines are set by a combined
effect of limb brightening and the different emission level between
coronal holes and quiet Sun.

\end{abstract}


\section{Introduction}

As the strongest emission line in the vacuum ultraviolet (VUV)
spectral range, the hydrogen Lyman-alpha (\lya) plays a crucial role
in the radiative energy transport in the lower transition region
(TR) \citep{Fontenla88}. Its radiance and profile provide important
information on the structure of the TR, where the solar wind flows
out through coronal funnels \citep{Tu05,Esser05}. The solar
\lya~line is also very important for interplanetary studies because
the spectral irradiance at the center of its profile is the main
excitation source responsible for the atomic hydrogen resonant
scattering in cool cometary and planetary materials
\citep{Emerich05}.

At the times of {\it Skylab} \citep[e.g.,][]{Nicolas76} and {\it OSO
8} (Orbiting Solar Observatory)
\citep[e.g.,][]{Lemaire78,Kneer81,Vial82,Bocchialini96}, full
\lya~and \lyb~line profiles were obtained. \lya~spectra in different
locations of the Sun were also obtained by the HRTS (High Resolution
Telescope and Spectrograph) instrument on rocket flights
\citep{Basri79} and by the UVSP instrument onboard {\it SMM}
\citep[Solar Maximum Mission,][]{Fontenla88}. These early
observations provided valuable information on the Lyman line
profiles. However, since these observations were made in Earth
orbits, the obtained Lyman line profiles were hampered by the
geocoronal absorption at the center.

The SUMER \citep[Solar Ultraviolet Measurements of Emitted
Radiation,][]{Wilhelm95,Lemaire97} observations at the first
Lagrangian point overcame this problem. The whole hydrogen Lyman
series is covered by the SUMER spectral range. By analyzing SUMER
spectra, \citet{Warren98} found that the average profiles for
\lyb~through {Ly$-\epsilon$\,($n$=5)} are self-reversed and show a
strong enhancement in the red wings. They also found that the peak
separation of these line profiles is larger at limb than at disk
center. \citet{Xia03} found that the asymmetry of the average
\lyb~line profile - the red-peak dominance - is stronger in the
quiet Sun than in a coronal hole. Also in sunspots SUMER observed
Lyman line profiles ($n\geq$2) and \cite{Tian09a} found that they
are almost not reversed, indicating a much smaller opacity above
sunspots than in surrounding regions. However, since the \lya~line
is so prominent, its high radiance leads to a saturation of the
detector microchannel plates. Although attempts were made to observe
\lya~on the bare part of the detector, the signal determination was
highly uncertain due to the gain-depression correction
\citep{Teriaca5a,Teriaca5b,Teriaca06}.

High-quality \lya~profiles without geocoronal absorption were
obtained after June 2008, when several non-routine observations were
made by SUMER. By closing the aperture door to reduce the incoming
photon flux to a level of about 20\%, the full \lya~profiles were
obtained \citep{Curdt08, Tian09b}. It turned out that the average
\lya~profile in the quiet Sun is strongly reversed and has a
stronger blue peak. Moreover, this asymmetry is stronger in regions
where the downflows are stronger in the TR.

Here we present new results from these unique data sets. Emphasis is
put on the results from a more recent observation in a polar coronal
hole region. The different behaviors of \lya~and \lyb~ profiles in
the coronal hole as compared to the quiet Sun are presented and
discussed. The data set allows for a study of the ratio between
\lya~and \lyb~radiances in the coronal hole. We also investigate the
limb brightening effect and the different radiances between coronal
holes and quiet Sun for the two Lyman lines as well as Si\,{\sc iii}
and O\,{\sc vi}, by studying their radiance distributions.

\section{Observations}

\begin{table}[]
\caption[]{ Emission lines used in this study. Here $\lambda$ and
$T$ represent the rest wavelength and formation temperature,
respectively.} \label{table1}
\begin{center}
\begin{tabular}{p{1.5cm} p{1.2cm} p{1.5cm}| p{1.0cm} p{1.1cm} p{1.5cm}}
\hline Ion & $\lambda$ ({\AA}) &  $\log(T/\rm{K})$
     & Ion & $\lambda$ ({\AA}) &  $\log(T/\rm{K})$ \\
\hline H\,{\sc i}~\lya & 1215.67 & 4.0 & Si\,{\sc iii} & 1206.51 & 4.7\\
H\,{\sc i}~\lyb & 1025.72 & 4.0 & O\,{\sc vi} & 1031.93 & 5.5\\
\hline
\end{tabular}
\end{center}
\end{table}

As mentioned in \citet{Curdt08}, we scanned six quiet-Sun regions
with a size of $120^{\prime\prime}\times120^{\prime\prime}$ at
different locations in the equatorial plane on 24 and 25 June 2008.
Three regions along the central meridian and including the southern
polar region were scanned on 26 June 2008. For these scans, profiles
of \lya~and Si\,{\sc iii} ($\lambda$\,1206~{\AA}) lines were
transmitted to the ground. The scanned region in the southern polar
region is outlined in white and is superposed on an XRT \citep[X-ray
Telescope,][]{Golub07} image, as shown in the right panel of
Fig.~\ref{fig.1}. On 23 September 2008, we added a second wavelength
setting for \lyb~and O\,{\sc vi} ($\lambda$\,1032~{\AA}) and scanned
a quiet-Sun region at disk center \citep{Tian09b}. The rest
wavelengths and formation temperatures of the four lines are listed
in Table~\ref{table1}.

More recently, we adopted these two wavelength settings and scanned
a region inside a large coronal hole in the southern polar region
from 16:01 to 17:19 on 17 April 2009. Similarly to previous
observations, we partly closed the aperture door, and could thus
reduce the input photon rate by a factor of $\approx5$. As a
prologue to the observation, full-detector images in the Lyman
continuum around {880~\AA} were obtained with open and
partially-closed door. In this way, accurate values of the photon
flux reduction could be established. After this prologue, the slit~7
(0.3\arcsec~$\times$~120{\arcsec}) was used to scan the target with
a size of about 150\arcsec~$\times$~120{\arcsec}, with an exposure
time of 15~s. The scanned region is outlined in white and is
superposed on an XRT image, as shown in the middle panel of
Fig.~\ref{fig.1}.

The standard procedures for correcting and calibrating the SUMER
data were applied, including local-gain correction, dead-time
correction, flat-field correction, destretching, and radiometric
calibration. Finally, the radiances of the spectra were divided by
the factor of the photon flux reduction, which was 18.4\% in this
observation.

\section{Results and discussions}
\subsection{Peak separation of \lya}
Due to the radiative transfer effect, a central reversal and two
peaks in the wings are normally present in \lya~profiles. The peak
separation can be regarded as an indicator of the opacity. It has
been found that the peak separation is larger at limb positions than
at disk center \citep{Curdt08}. Here we aim at finding possible
differences in the peak separation between coronal holes and
quiet-Sun regions at the limb.

\begin{figure}
\centering {\includegraphics[width=\textwidth]{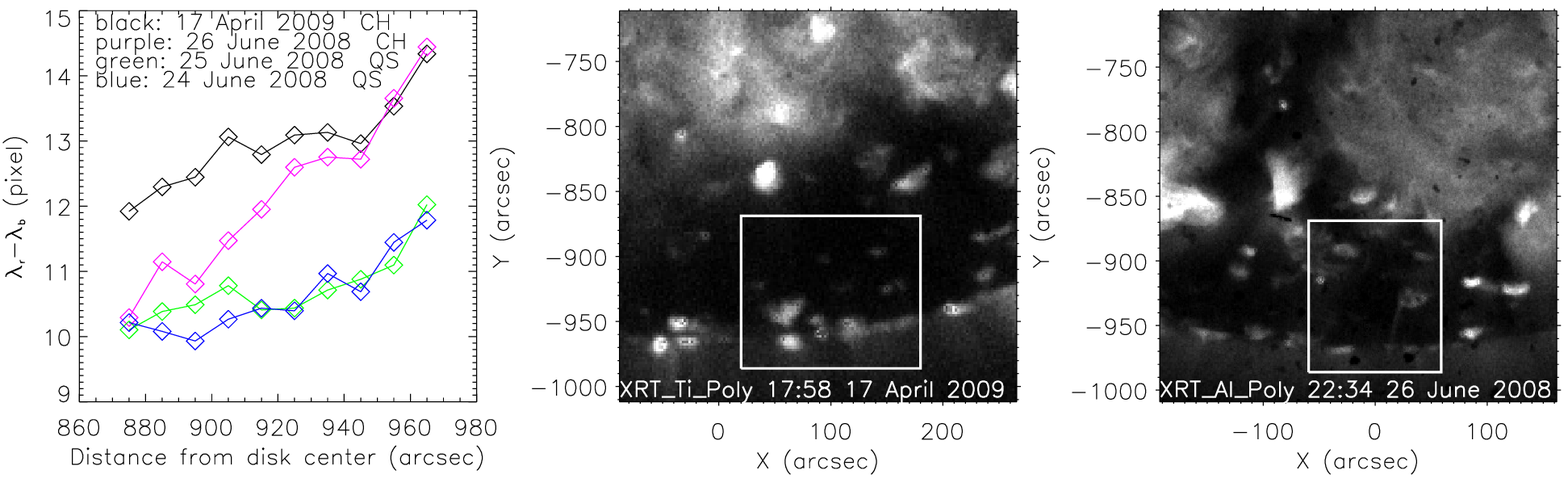}} \caption{
Left: the variation of peak separation with the distance from disk
center. Middle: XRT image taken at 17:58 on 17 April 2009. Right:
XRT image taken at 22:34 on 26 June 2008. In the middle and right
panels, the images are displayed in the linear scale; and the white
rectangles outline the coronal hole regions scanned by SUMER. }
\label{fig.1}
\end{figure}

We selected four limb scans for this study. For each scan, we first
sorted all the data points by the distance from disk center and
defined 10 bins. The \lya~profiles in each bin were then averaged to
pick up general properties out of the large solar variability. The
signal to noise ratio is, consequently, very high. By applying a
second-order polynomial fit to both peaks of the average profile, we
determined the spectral positions of the two peaks $\lambda_b$ and
$\lambda_r$. The variations of the peak separation with the distance
from disk center are shown in the left panel of Fig.~\ref{fig.1}.
The off-disk profiles were excluded in the above calculations since
their shapes approximate a Gaussian and show no peaks in the wings.

From Fig.~\ref{fig.1} we find that in both coronal holes and the
quiet Sun the peak separation of the \lya~profile increases towards
the limb, indicating a larger opacity combined with a source
function decreasing with altitude. This result complements the
finding by \citet{Warren98}, in which the authors found that the
peak separations of the average profiles for \lyb~through
{Ly$-\epsilon$\,($n$=5)} are larger at limb than at disk center.

It is also obvious that the peak separation in the polar coronal
hole is larger than those in the quiet-Sun region at the east limb.
The different behavior of the two coronal holes can be understood if
we check the positions of the scans with respect to the coronal hole
boundaries. From Fig.~\ref{fig.1} it is clear that the scanned
region on 17 April 2009 was well inside the polar coronal hole.
However, the upper part of the scanned region on 26 June 2008 was
very close to the boundary, so that the radiance there was likely to
be contaminated by the nearby quiet-Sun emission. This effect might
explain the fact that the peak separation increases from the
quiet-Sun level to the coronal hole level, for the observation on 26
June 2008. Another possibility might be that although coronal
radiation is much reduced in the coronal hole (see the next
paragraph), from the nearby quiet-Sun structures we may still get
quiet a bit of radiation flux which ionizes hydrogen atoms and
reduces the opacity. This effect should be weaker with increasing
distance from the boundary.

This result indicates a larger opacity in coronal holes, as compared
to the quiet Sun. There might be two possible explanations. First,
the magnetic-field lines in polar coronal holes are almost
perpendicular to the line of sight, whilst they are aligned in
various directions at the east limb. It has been shown that since
the variations of density and temperature are different when seen
across and along the magnetic field lines, the self-reversal of the
Lyman line profiles in a prominence can be different if the
prominence is observed from different viewing angles
\citep{Heinzel05,Schmieder07}. In coronal holes and quiet-Sun
regions, we should not exclude the possibility that the different
magnetic structures might influence the \lya~profile in the
processes of emission and absorption. It is also possible that the
larger opacity is the result of a weaker radiation field in the
upper atmosphere of coronal holes. Due to the lower radiative
ionization, more atomic hydrogen will be populated in the upper TR
and corona, which leads to a stronger absorption of the profiles. In
order to fully understand this phenomenon, sophisticated models
including calculations of non-LTE (local thermodynamic equilibrium)
radiative transfer should be developed in the future.

\subsection{Asymmetry of Lyman line profiles}
Previous observations have shown that the dominant asymmetry for
profiles of the hydrogen \lya~and higher order Lyman lines is
opposite in the quiet Sun. Most \lya~profiles have a stronger blue
peak \citep{Curdt08, Tian09b}, while most \lyb~profiles have a
stronger red peak \citep{Warren98, Xia03}. The asymmetries are
probably produced by the combined effects of the differential flows
in the solar atmosphere and different line opacities
\citep{Gouttebroze78, Fontenla02, Gunar08,Curdt08, Tian09b}.

\begin{figure}
\centering {\includegraphics[width=\textwidth]{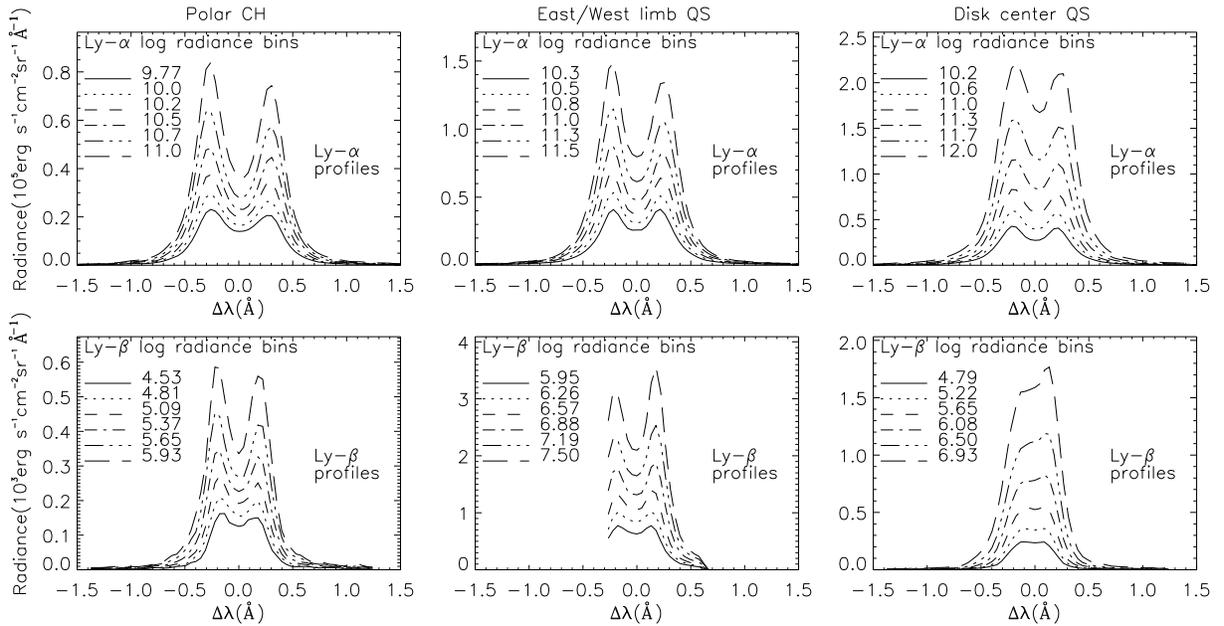}} \caption{
Averaged \lya~(upper panels) and \lyb~(lower panels) profiles in six
different radiance bins, as obtained from a polar coronal hole (left
panels, 17 April 2009), a quiet-Sun region at disk center (right
panels, 23 September 2008), and quiet-Sun regions at east limb
(upper middle panel, 25 August 2008) and west limb (lower middle
panel, 07 June 1996). The level of each bin is also shown in the
plots. Here we have excluded spectra acquired off-disk. }
\label{fig.2}
\end{figure}

Similar to \cite{Curdt08}, here we present in Fig.~\ref{fig.2} the
average profiles of \lya~and \lyb~in six bins which are equally
spaced in radiance. The off-disk profiles were excluded from the
averaging. The \lyb~profiles at the west limb were acquired from
13:25 to 14:36 on 07 June 1996, with an exposure time of 10~s. For
this observation only 25 spectral pixels were recorded so that the
profiles were not complete. Profiles presented in the other five
panels are all from the non-routine observations mentioned above.
From Fig.~\ref{fig.2} it is clear that the Lyman line profiles are
more reversed at limb positions than at disk center, which has
already been found by \citet{Warren98} and \citet{Curdt08}.

The most interesting feature in Fig.~\ref{fig.2} is that the
\lyb~profiles in the polar coronal hole have an asymmetry opposite
to those in the quiet Sun, while the dominant asymmetry of the
\lya~profiles is the same in different locations of the Sun. We
noticed that \citet{Xia03} found more locations with blue-peak
dominance in \lyb~profiles in equatorial coronal holes than in
quiet-Sun regions. However, the average \lyb~profile in
\citet{Xia03} is still stronger in the red peak. Here we find very
clearly that most \lyb~profiles are stronger in the blue peak in the
polar coronal hole.

Since the asymmetries of the Lyman line profiles are likely to be
influenced by flows in various layers of the solar atmosphere, and
we know that upflows are predominant in the upper TR and lower
corona in coronal holes \citep{Dammasch99,Hassler99,Tu05}, it is
natural to relate the upflows to the profile asymmetries. In order
to investigate this relationship, newly designed observations will
be done in the near future. However, the flows of the emitting
material might also be different in the coronal hole and quiet Sun,
which will also alter the asymmetries of line profiles.

\subsection{Radiance ratio between \lya~and \lyb}

The \lya/\lyb~ratio is very sensitive to the physical and
geometrical properties of the fine structures in prominences
\citep{Vial07}. However, this ratio in coronal holes is not well
established.

\begin{figure}
\centering {\includegraphics[width=0.6\textwidth]{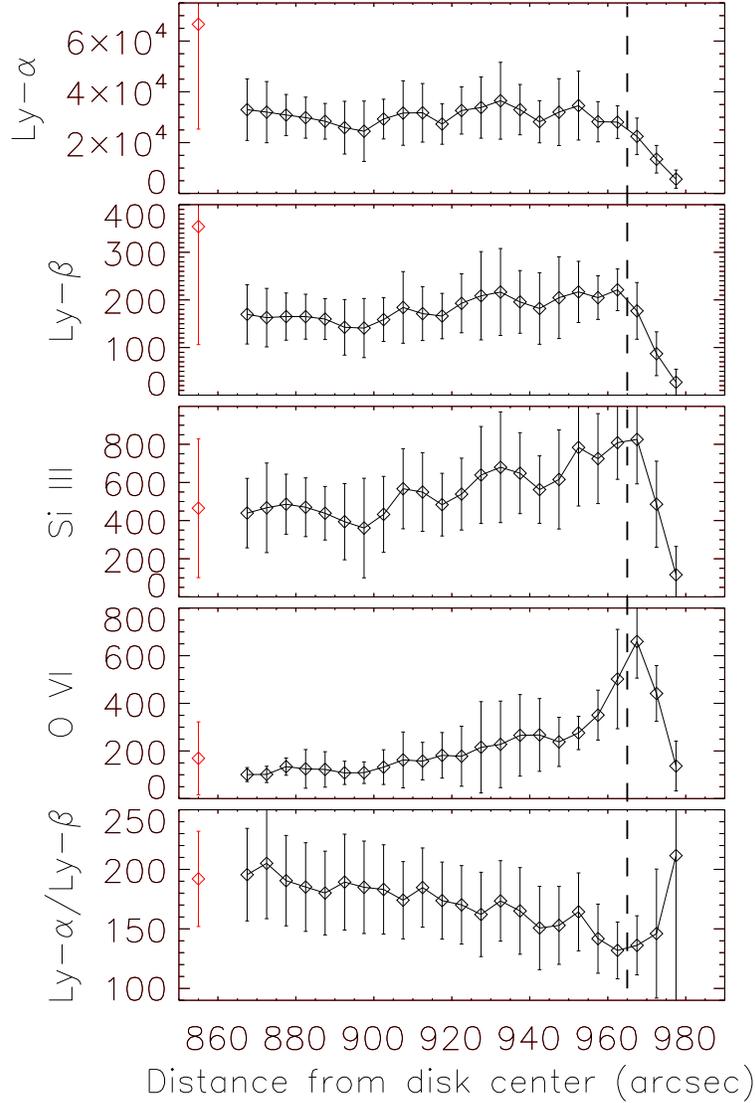}} \caption{
The radiances of \lya, \lyb, Si\,{\sc iii}, and O\,{\sc vi}, and the
radiance ratio of \lya~and \lyb, as obtained from the polar coronal
hole on 17 April 2009, are shown as a function of the distance from
disk center. In each bin, the diamond and the vertical bar represent
the median value and standard deviation, respectively. The median
values and the corresponding standard deviations obtained from the
disk-center quiet-Sun region on 23 September 2008 are also shown and
marked in red for comparison. The unit of the radiance is
$\rm{erg~cm^{-2}~s^{-1}~sr^{-1}}$. The dashed line indicates the
approximate position of the limb. } \label{fig.3}
\end{figure}

The spectra obtained on 17 April 2009 include both \lya~and
\lyb~lines, and, thus, allow for studying the center to limb
variation of the radiance ratio between \lya~and \lyb~in the coronal
hole. In Fig.~\ref{fig.3} we present the radiances of four lines and
the ratio \lya/\lyb~ in 23 bins which are equally spaced in
distance. In each bin, the diamond and the vertical bar represent
the median value and standard deviation, respectively. The median
values and the corresponding standard deviations obtained from the
disk-center quiet-Sun region on 23 September 2008 are also shown and
marked in red for comparison. The radiances are given in energy unit
of $\rm{erg~cm^{-2}~s^{-1}~sr^{-1}}$.

From Fig.~\ref{fig.3}, we can see that the effect of limb
brightening is obviously present in the line radiance of O\,{\sc
vi}, and also clear in Si\,{\sc iii}. This effect is totally absent
in \lya, which is consistent with \cite{Curdt08}. The effect is
present, although not prominent, for \lyb. This is because \lyb~has
a smaller opacity and thus should behave more similarly to optically
thin lines than \lya. As a result, the ratio \lya/\lyb~decreases
towards the limb, from the quiet-Sun level of about 190 to 130 at
the limb.

Above the limb, the ratio increases further because the opacity of
\lya~stays higher than one while the lower-than-one \lyb~opacity
leads to a lower radiance. An additional effect might result from
the different conditions of formation of the two lines when one goes
higher in the atmosphere: the \lya~line becomes scattering dominated
(and proportional to the density) while the \lyb~line stays
collision-dominated (and proportional to the square of the density).

\subsection{Radiance distribution}

Radiance distributions of EUV lines have been intensively studied
\citep[e.g.,][]{Stucki02,Raju06,Pauluhn07}. Radiance histograms of
coronal holes are shifted towards lower values, having a narrower
and higher peak indicative of more uniform radiances, as compared to
those of quiet-Sun regions. As shown in \citet{Raju06}, the radiance
distributions are more similar in the quiet-Sun region and the
coronal hole for lower-TR lines, while they become increasingly
different for lines formed in the upper-TR and corona.

\begin{figure}
\centering {\includegraphics[width=0.8\textwidth]{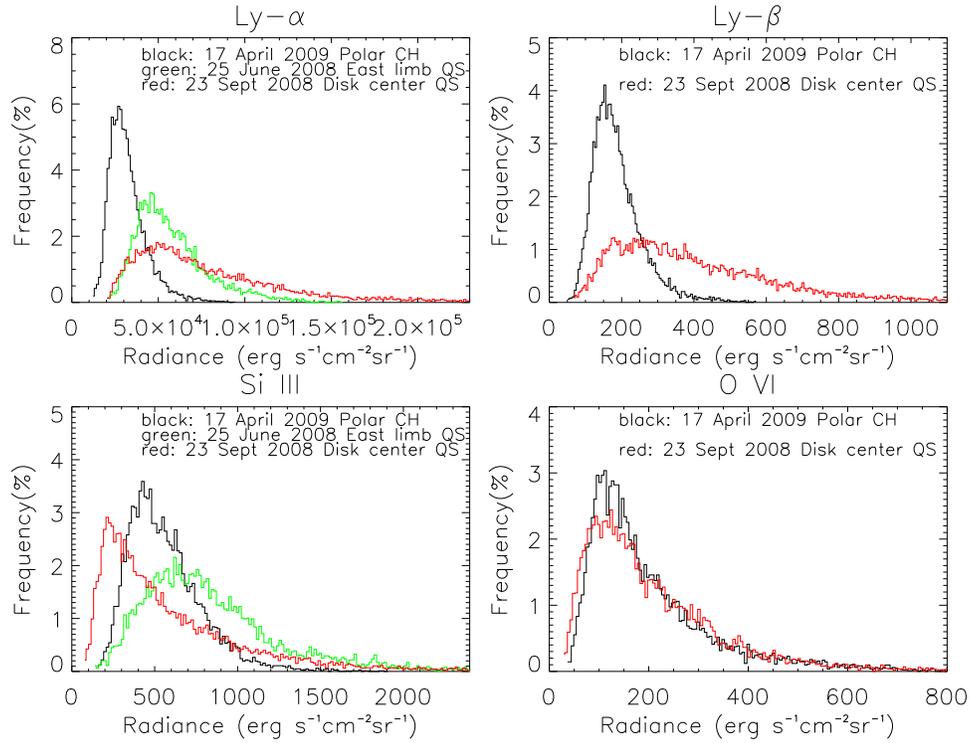}} \caption{
Radiance histograms of \lya~(upper left), \lyb~(upper right),
Si\,{\sc iii}~(lower left), and O\,{\sc vi}~(lower right) at
different locations of the Sun. Spectra obtained off-disk were
excluded in the analysis. } \label{fig.4}
\end{figure}

In Fig.~\ref{fig.4} we present the radiance histograms of \lya,
\lyb, Si\,{\sc iii}, and O\,{\sc vi} at different locations of the
Sun, by using the spectra obtained in the non-routine observations
mentioned above. Again, off-disk data points were excluded from the
histograms.

At first sight, the behaviors of the radiance distributions of
Si\,{\sc iii} and O\,{\sc vi} are "peculiar". According to
\cite{Raju06}, the radiance distribution of the lower-TR line
Si\,{\sc iii} should be similar in coronal holes and in the quiet
Sun, and the radiance distribution of the upper-TR line O\,{\sc vi}
should be shifted towards the weaker side in coronal holes. However,
we have to bear in mind that the east-limb region and the polar
coronal hole in our study are much closer to the limb, as compared
to the polar regions in \cite{Raju06}. Thus, the limb brightening
effect is much more prominent in our distributions. Due to this
effect, the radiance of Si\,{\sc iii} should be weaker at disk
center, and stronger at east/west limb. In polar coronal holes, the
radiance should also be stronger than that at disk center, and
perhaps slightly weaker than that at east/west limb. This is exactly
what we see in the radiance distributions of Si\,{\sc iii}. For
O\,{\sc vi}, a combined effect of dark coronal hole emission and
limb brightening makes its radiance distribution similar in the
polar hole and at disk center. These two effects are clearly
revealed by the median of O\,{\sc vi} radiances, as shown in
Fig.~\ref{fig.3}.

The median values of the radiances shown in Fig.~\ref{fig.3} also
reveal that the polar coronal hole is darker in \lya~and \lyb,
although these two lines are formed in the lower TR. This behavior
is likely due to the large optical thickness. The reduced ionizing
coronal radiation may also play a role, as in the case of the
He\,{\sc ii} line \citep[e.g.][]{Raju06}. Since the lines of
\lya~and \lyb~have no or little limb brightening, their radiance
distributions in coronal holes are shifted towards the weaker side,
as shown in Fig.~\ref{fig.4}.

\section{Summary}
Full hydrogen \lya~ profiles which are clean from geocoronal
absorption were acquired by SUMER with high spectral and spatial
resolutions, through several non-routine observations. In some of
these observations, in addition \lyb, Si\,{\sc iii}, and O\,{\sc vi}
profiles were recorded (quasi-) simultaneously.

The peak separations of \lya~profiles are found to be larger in
coronal holes than in the quiet Sun, indicating a larger opacity in
coronal holes. This difference might be due the different magnetic
structures or the different radiation fields in the two regions. We
also found that the dominant asymmetry of the \lyb~profiles in the
polar coronal hole is opposite to that in the quiet Sun. In order to
understand this phenomenon, we need to investigate the influence of
the upflows in the upper TR and the flows of the emitting materials
on the line profiles. We also investigated the center to limb
variation of the radiance ratio between \lya~and \lyb~, which has a
declining trend towards the limb in the coronal hole. Finally, the
radiance distributions of the four lines were explained by taking
into account the effects of limb brightening and darker emission in
coronal holes.

In order to fully understand the above new results, non-LTE models
including detailed calculations of radiative transfer should be
developed in the future.

\begin{acknowledgements}
{\bf Acknowledgements:}  SUMER, which is financially supported by
DLR, CNES, NASA, and the ESA PRODEX Programme (Swiss contribution),
is an instrument onboard {\it SOHO}. XRT is an instrument onboard
{\it Hinode}, a Japanese mission developed and launched by
ISAS/JAXA, with NAOJ as domestic partner and NASA and STFC (UK) as
international partners. It is operated by these agencies in
co-operation with ESA and NSC (Norway). Hui Tian is supported by the
IMPRS graduate school run jointly by the Max Planck Society and the
Universities of G\"ottingen and Braunschweig. The work of Hui Tian's
group at Peking University is supported by NSFC under contract
40874090.
\end{acknowledgements}

\end{document}